\documentstyle[11pt,aaspp4]{article}

\lefthead{}
\righthead{}

\begin{document}

\title{Numerical Simulations of Convective Accretion Flows \\
       in Three Dimensions}

\author{Igor V. Igumenshchev\altaffilmark{1,2}, 
        Marek A. Abramowicz\altaffilmark{1,3}, and
        Ramesh Narayan\altaffilmark{4}}



\altaffiltext{1}{Institute for Theoretical Physics, G\"oteborg University and 
Chalmers University of Technology, 412 96 G\"oteborg, Sweden;
ivi@fy.chalmers.se, marek@fy.chalmers.se.}
\altaffiltext{2}{Institute of Astronomy, 48 Pyatnitskaya Street, 
109017 Moscow, Russia.}
\altaffiltext{3}{Laboratorio Interdisciplinare SISSA, Trieste, Italy, and ICTP,
Trieste, Italy.}
\altaffiltext{4}{Harvard-Smithsonian Center for Astrophysics, 60 Garden Street,
Cambridge MA 02138, U. S. A.; narayan@cfa.harvard.edu.}

\begin{abstract}

Advection-dominated accretion flows (ADAFs) are known to be
convectively unstable for low values of the viscosity parameter
$\alpha$.  Two-dimensional axisymmetric hydrodynamic simulations of
such flows reveal a radial density profile which is significantly
flatter than the $\rho\propto r^{-3/2}$ expected for a canonical ADAF.
The modified density profile is the result of inward transport of
angular momentum by axisymmetric convective eddies.  We present
three-dimensional hydrodynamic simulations of convective ADAFs which
are free of the assumption of axisymmetry.  We find that the results
are qualitatively and quantitatively similar to those obtained with
two-dimensional simulations.  In particular, the convective eddies are
nearly axisymmetric and transport angular momentum inward.

\end{abstract}


\section{Introduction}

Analytical and numerical calculations have shown that
advection-dominated accretion flows (ADAFs) experience strong
convection, especially for low values of the viscosity parameter
$\alpha$ (Narayan \& Yi 1994, 1995; Igumenshchev, Chen \& Abramowicz
1996; Igumenshchev \& Abramowicz 1999, 2000; Stone, Pringle \&
Begelman 1999).  The numerical work reveals that convection
significantly modifies the structure of ADAFs (Stone et al. 1999;
Narayan, Igumenshchev \& Abramowicz 2000, hereafter NIA).  The change
in the structure is understood to be a consequence of inward transport
of angular momentum by convective eddies (NIA; Quataert \& Gruzinov
2000, hereafter QG; Ryu \& Goodman 1992; Stone \& Balbus 1996).

All the numerical simulations done so far have corresponded to
axisymmetric two-dimensional flows.  NIA and QG emphasized that the
inward transport of angular momentum seen in these simulations is a
consequence of axisymmetry (cf. Stone \& Balbus 1996) and that the
results could in principle be qualitatively very different in three
dimensions.  We present in this paper the first fully
three-dimensional numerical simulations of ADAFs.  The details of our
numerical technique are given in \S2 and the results are described in
\S3.  The paper concludes with a discussion in \S4.

\section{Numerical Method}

We solve the non-relativistic time-dependent Navier-Stokes equations
of an accretion flow in the gravitational potential of a central black
hole:
$$
{d \rho\over dt}+\rho\nabla\cdot\vec{v}=0, \eqno(2.1)
$$
$$
\rho{d \vec{v}\over dt}=-\nabla P + \rho\nabla\Phi + \nabla{\bf \Pi},
\eqno(2.2)
$$
$$
\rho{d e \over dt}=-P\nabla\cdot\vec{v}+Q, \eqno(2.3)
$$
where $\rho$ is the density, $\vec{v}$ is the velocity, $P$ is the
pressure, $\Phi=-GM/r$ is the Newtonian potential of the central mass $M$, 
$e$ is the specific internal energy, ${\bf \Pi}$ is the viscous
stress tensor (we include all components), and $Q$ is the viscous
dissipation function.  Since we assume an ADAF, there is no radiative
cooling.  We adopt the ideal gas equation of state, $P=(\gamma-1)\rho
e$, where $\gamma$ is the adiabatic index, and write the kinematic
shear viscosity coefficient as $ \nu=\alpha{c_s^2/\Omega_K}$, where
$c_s=(P/\rho)^{1/2}$ is the isothermal sound speed and
$\Omega_K=(GM/r^3)^{1/2}$ is the Keplerian angular velocity.
Equations (2.1)--(2.3) are independent of the black hole mass $M$ when
the radius $r$ is scaled by the gravitational radius $r_g=2GM/c^2$ and
time is scaled by $r_g/c$.

Each time step of the numerical computation is split into two
sub-steps: hydrodynamical and viscous. The hydrodynamical sub-step is
calculated by using the PPM algorithm (Colella \& Woodward 1984), and
the viscous sub-step is solved by using a standard explicit algorithm
for the viscous contribution to the equation of motion (2.2) and the
energy equation (2.3).

We solve equations (2.1)--(2.3) on a 3D Cartesian grid consisting of a
sequence of six nested sub-grids.  Each sub-grid $k$ is $26\times
26\times 52$ in size and is uniform, extending over the range
($0,R_k^{out}$), ($0,R_k^{out}$), ($-R_k^{out},R_k^{out}$) in $x$,
$y$, $z$, respectively.  The innermost grid has $R_1^{out}=26r_g$ and
each successive $R_k^{out}$ is equal to $2R_{k-1}^{out}$, such that
the outermost sub-grid extends to $R_6^{out}=832r_g$.  Figure~1
illustrates the geometry of the grid.  Due to numerical limitations we
compute the flow only over a quarter of the full cubic volume and
apply azimuthal periodic boundary conditions in the $(x,z)$ and
$(y,z)$ planes.  At the outer boundary $R_6^{out}$ we use a free
outflow boundary condition, and in the innermost sub-grid we assume
that matter drains freely inside a sphere of radius $r\simeq 3 r_g$.
The inner boundary condition mimics the effect of the last stable
orbit around a Schwarzschild black hole.

\begin{figure}
\plotone{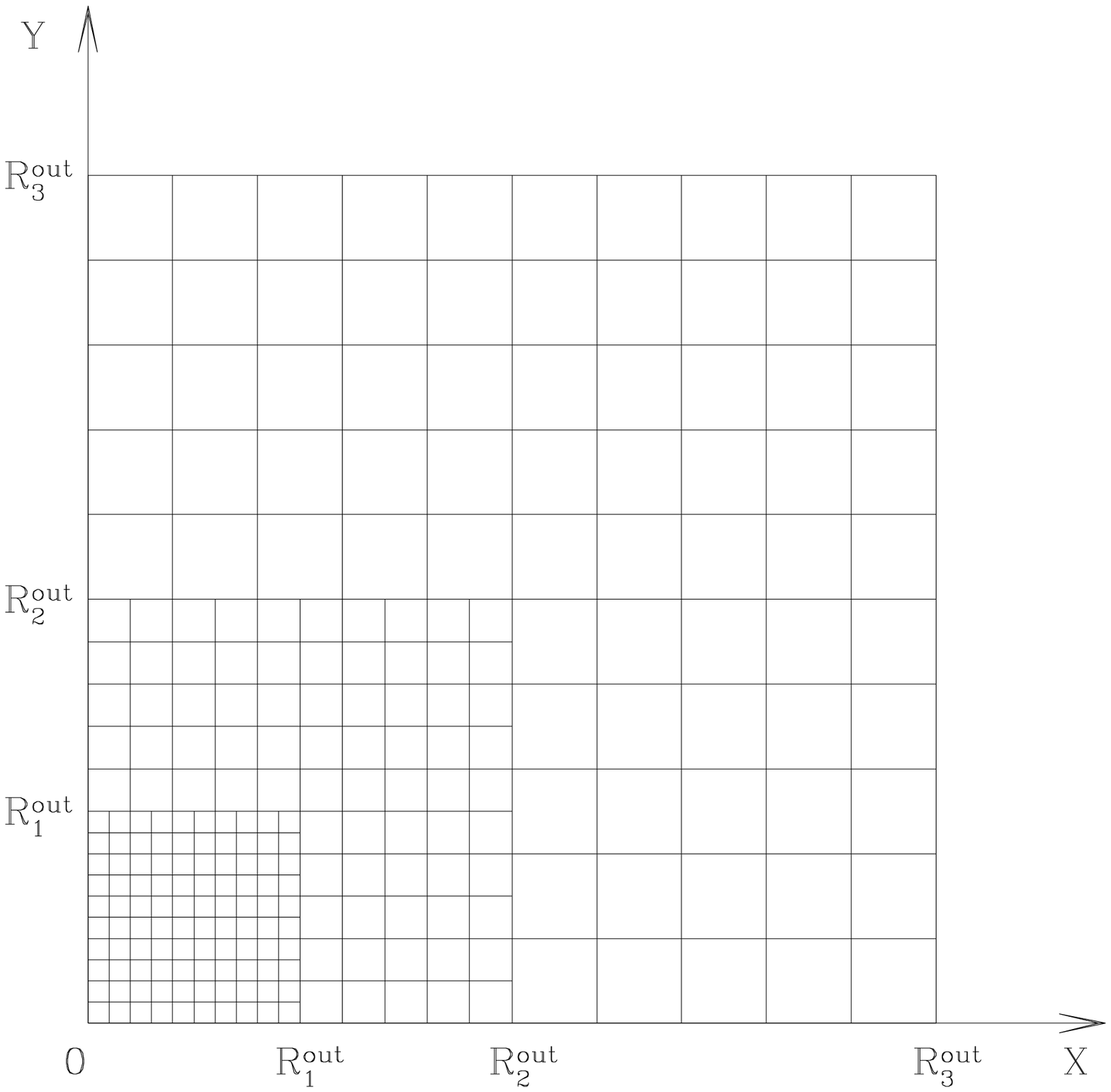}
\caption
{A 2D illustration of the nested Cartesian grid used in the
calculations.  The example shown has three sub-grids, each of size
$10\times10$.
\label{fig1}}
\end{figure}

\placefigure{fig1}

The time-step $\Delta t$ for the numerical integration is chosen in
accordance with the Courant condition for the hydrodynamical sub-step.
For this $\Delta t$, the viscous integration is numerically stable
because we choose sufficiently low values for the viscosity parameter
$\alpha$.  We assume that mass is injected into the computational
domain over an equatorial slender torus with radius $r_{inj}\approx
800 r_g$.  The matter is injected at a fixed rate $\dot{M}_{inj}$ and
has rotational velocity (with respect to the $z$-axis) equal to $0.95$
times the local Keplerian velocity.  As a result of viscous
interactions, part of the injected matter moves inwards and forms an
accretion flow, while the rest leaves the computational domain through
the outer boundary.  We follow the time evolution of the flow on time
scales longer than the viscous time scale at $r_{inj}$.  This allows
the flow to attain steady state in a time-averaged sense.  Note
that when $\rho$ is scaled by $\dot{M}_{inj}$, the results are
independent of the value of $\dot{M}_{inj}$.  Therefore, the
simulations depend on only two dimensionless parameters: $\alpha$,
$\gamma$.

\section{Results}

We have calculated four models (A, B, C, D) which differ in the choice
of $\alpha$ and $\gamma$ (see Table~1).  All the models show
complicated time-dependent convective motions, analogous to those
described in detail by Igumenshchev et al. (1996), Igumenshchev \&
Abramowicz (1999, 2000) and Stone et al. (1999).  The amplitude of the
variability is not large, and the average properties of the flow are
well represented by time-averaging.  To good accuracy, the
time-averaged flow is axisymmetric in all the models.

\placetable{tbl-1}

Figure~2 shows the time-averaged radial profiles in the equatorial
plane ($z=0$) of $\rho$, $\Omega$ and $c_s$ for the four models.  In
the inner regions, $r\la 10^2 r_g$, the profiles are well approximated
by power-laws: $\Omega\propto r^{-3/2}$, $c_s\propto r^{-1/2}$,
$\rho\propto r^{-\beta}$, where the index $\beta$ is $\approx 0.5$ for
Models~A and C and $\approx 0.8$ for Models~B and D.  (For $r>2\times
10^2 r_g$, the flows are strongly affected by the outer boundary at $r
\approx 8\times 10^2 r_g$ and are not power-law in character.)  The
scalings of $\Omega$ and $c_s$ and that of $\rho$ in Models~A and C
are identical to those predicted by the analytical work of NIA and QG.
The $\beta\approx0.8$ scaling of Models~B and D is a little
surprising, and suggests that $\beta$ depends on the value of
$\gamma$.  A similar dependence of $\beta$ on $\gamma$ was found also
in the axisymmetric 2D simulations of Igumenshchev \& Abramowicz (2000).

\begin{figure}
\plotone{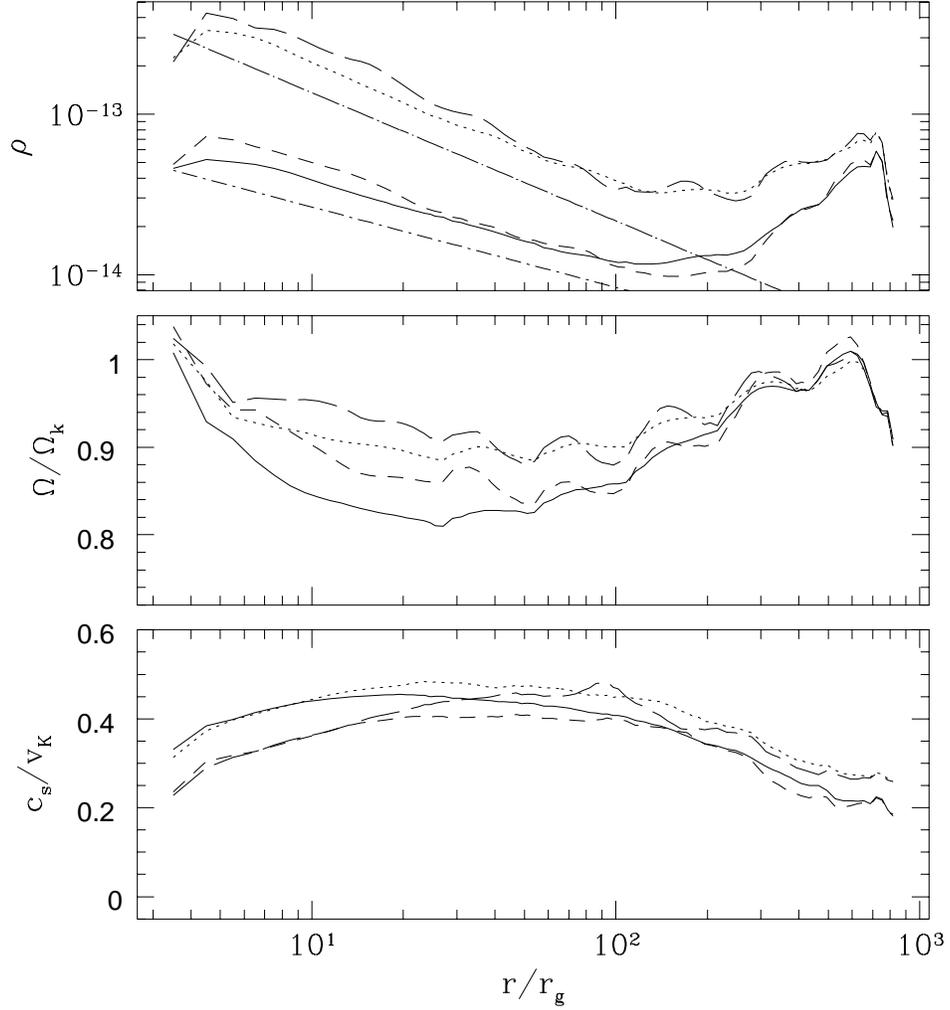}
\caption
{Radial profiles of the equatorial density $\rho$, angular velocity
$\Omega/\Omega_K$ and isothermal sound speed $c_s/r\Omega_K$ for
Models~A (solid lines), B (dotted lines), C (dashed lines) and D
(long-dashed lines).  The dot-dashed and dot-long-dashed lines in the
top panel show power-law profiles, $r^{-0.5}$ and $r^{-0.8}$, for
comparison.
\label{fig2}}
\end{figure}

\placefigure{fig2}

Figure 3 shows a snapshot of the flow in the equatorial plane for
Model C.  The other three models are qualitatively similar.  The most
important feature is that the density and velocity fluctuations are
almost axisymmetric, as seen by the fact that the eddies are
significantly elongated in the azimuthal direction.  This suggests
that the present 3D results are not likely to be very different from
those obtained with 2D axisymmetric simulations.  The minimum
length-scale of perturbations in the radial $r$ and polar $\theta$
directions depends on the viscosity parameter $\alpha$, being larger
for larger values of $\alpha$.

\begin{figure}
\plottwo{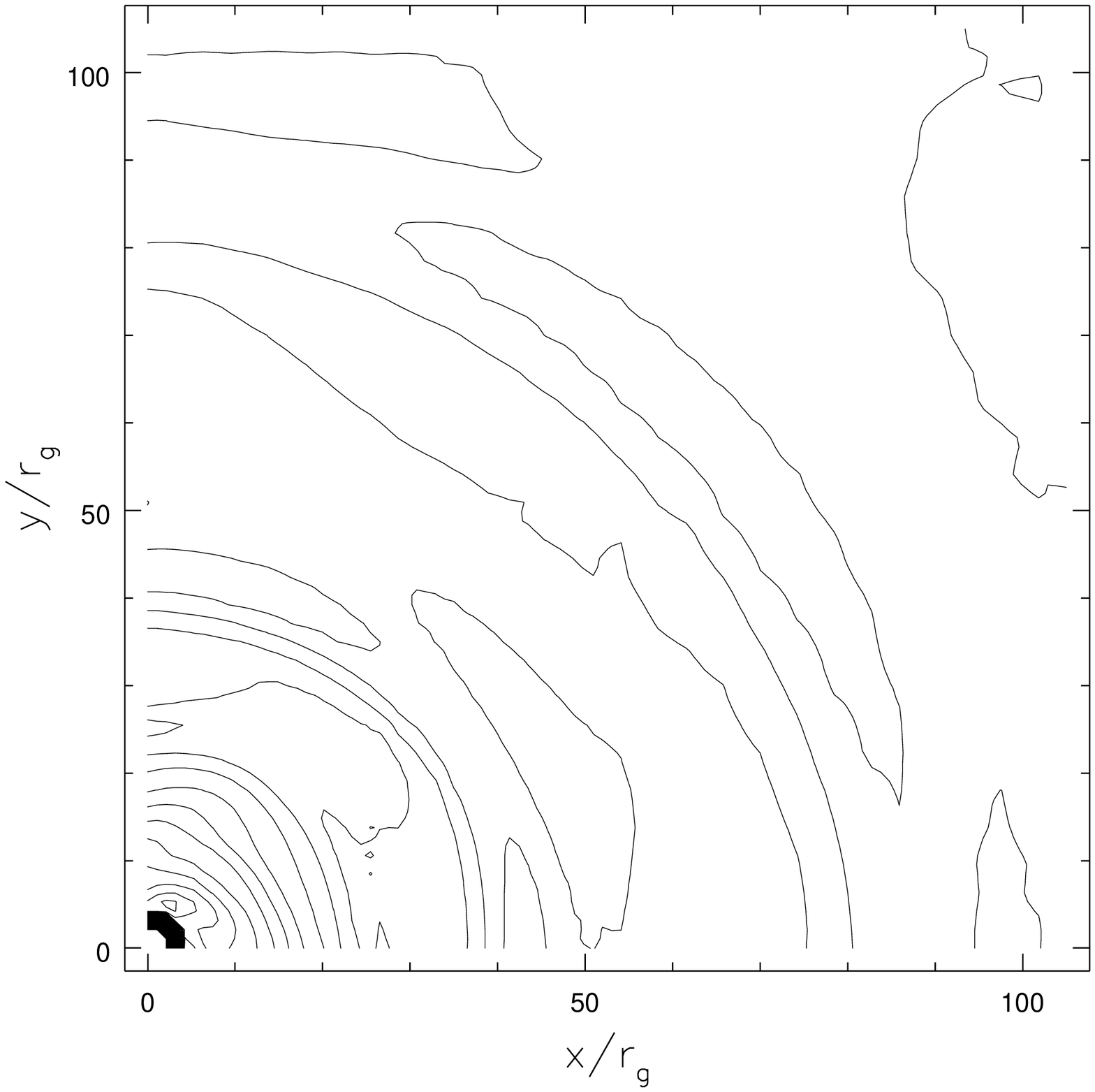}{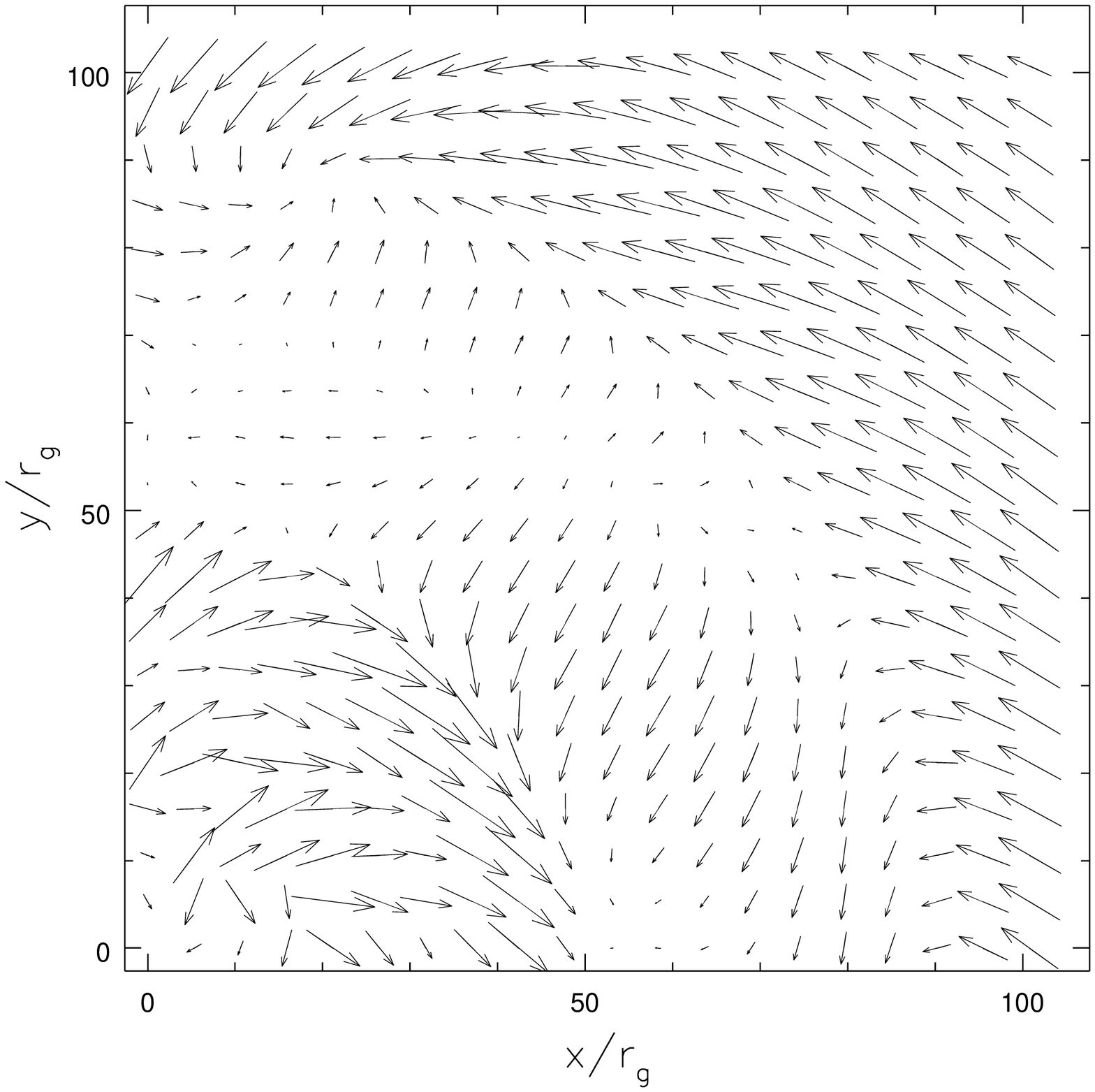}
\caption
{Snapshot view of the equatorial density (left) and velocity field
(right) in Model~C. The density contours are spaced with
$\Delta\log\rho=0.05$.  Arrows in the right panel show the
dimensionless relative velocity
$(\vec{v}-\langle\vec{v}\rangle)/|\langle\vec{v}\rangle|$, where
$\langle\vec{v}\rangle$ is the time-averaged velocity.  The background
rotation of the accretion flow is anti-clockwise.  Therefore, material
rotating clockwise/anti-clockwise in the figure has less/more specific
angular momentum than the background.
\label{fig3}}
\end{figure}

\placefigure{fig3}

The left panel in Fig.~4 shows a snapshot meridional cross-section of
Model~C in the $(y,z)$-plane.  The shading shows the distribution of
the vorticity vector, $\vec{\omega}={\rm rot}\vec{v}$.  Regions with a
positive $x$-component of the vorticity $\omega_x$ are shown in grey,
and regions with negative $\omega_x$ are shown in white.  It is seen
that the flow is dominated by radial streams of inflowing and
outflowing material which are sandwiched in the $\theta$-direction.
The streams are chaotic and move around so that they are washed out
when a time average is taken.  Models~A and B are very similar to
Model C, except that they have $2-3$ inflowing streams at a given
moment, whereas Model~C has usually $5-6$ streams.

\begin{figure}
\plottwo{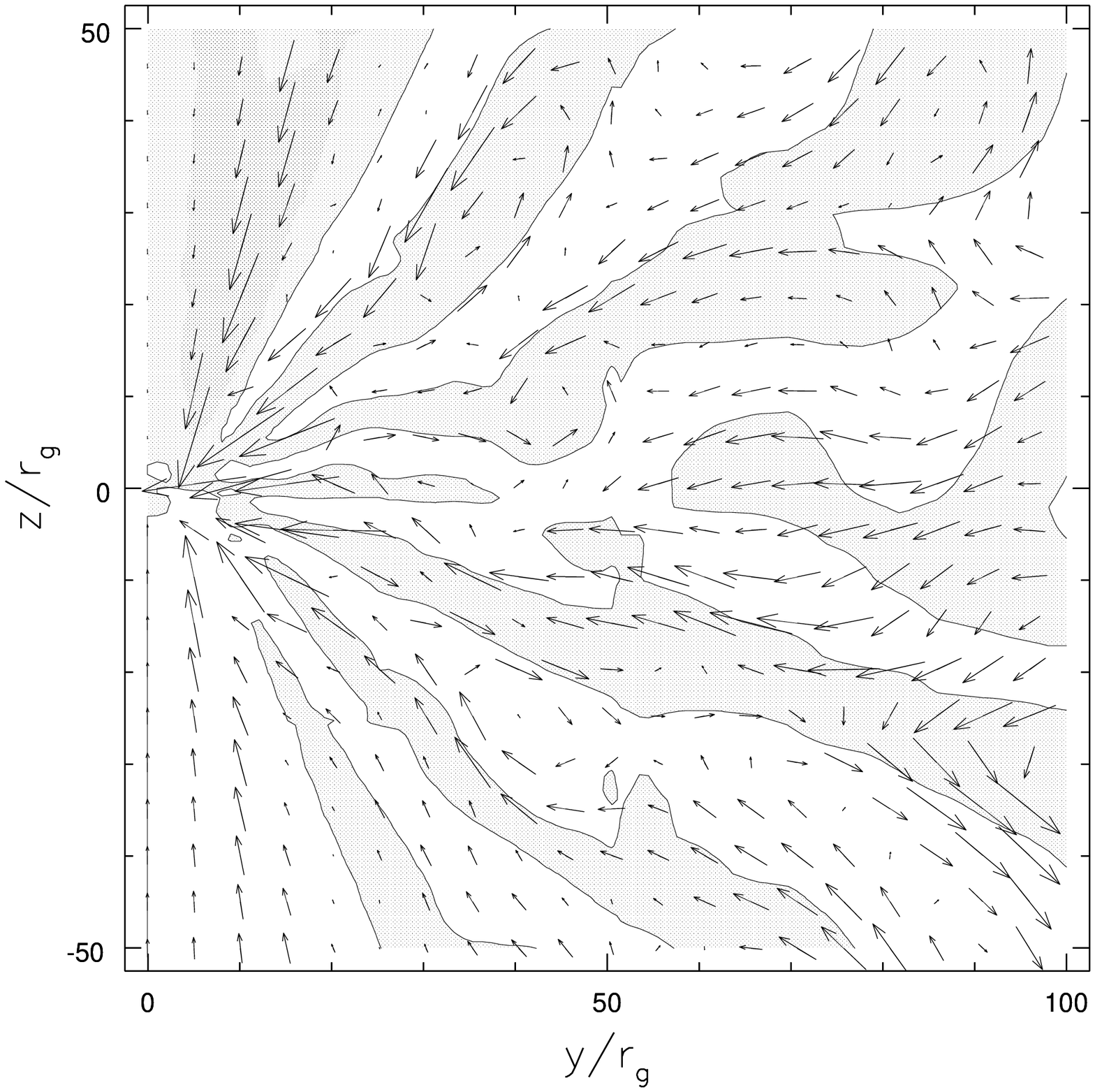}{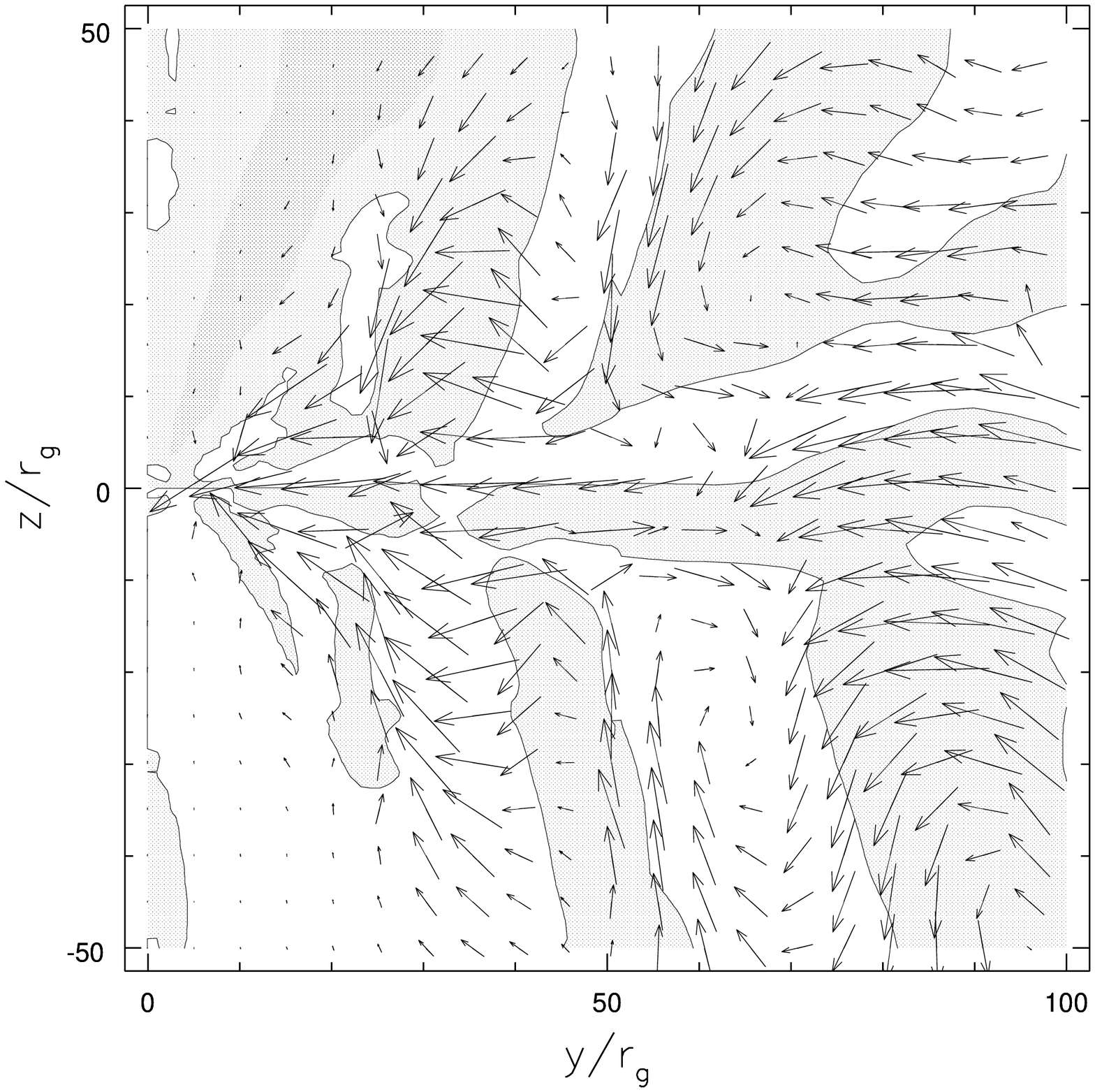}
\caption
{Vorticity and momentum vectors in the meridional $(y,z)$-plane
corresponding to a snapshot of Model~C (left) and a time-average of
Model~D (right).  Arrows correspond to $r\rho\vec{v}$.
\label{fig4}}
\end{figure}

\placefigure{fig4}

Model~D is qualitatively different from the other three models, as
shown by the right panel in Fig.~4.  What is plotted here is the {\it
time-averaged} flow pattern and vorticity.  We see that the model has
quasi-stationary structures which survive even after averaging
over about 100 periods of the Keplerian rotation at $r=100 r_g$.
The flow pattern is almost symmetrical with respect to the equatorial
plane, and shows the presence of a stable and regular arrangement of
circulation cells in the meridional plane.

The direction of transport of angular momentum is given by the sign of
the $(r\varphi)$-component of the Reynolds stress tensor, $\langle
v'_r v'_\varphi\rangle$.  NIA showed that this quantity is negative in
2D simulations of convective ADAFs, but noted that it was a
consequence of the axisymmetry of the simulations and that real 3D
convection might behave differently.  In Figure~5 we show the radial
dependence of the Reynolds stress averaged over $\theta$ for the four
3D models described in this paper.  We see that, except for a small
boundary region, the stress is negative over the bulk of the flow in
all the models.  This confirms that the 2D results published in
previous papers are physically meaningful.  Figure 5 shows that the
magnitude of the Reynolds stress is positively correlated with the
values of both $\alpha$ and $\gamma$.  Model~D has a surprisingly weak
stress.  This may be connected to the quasi-stationary meridional
circulation seen in this model.

\begin{figure}
\plotone{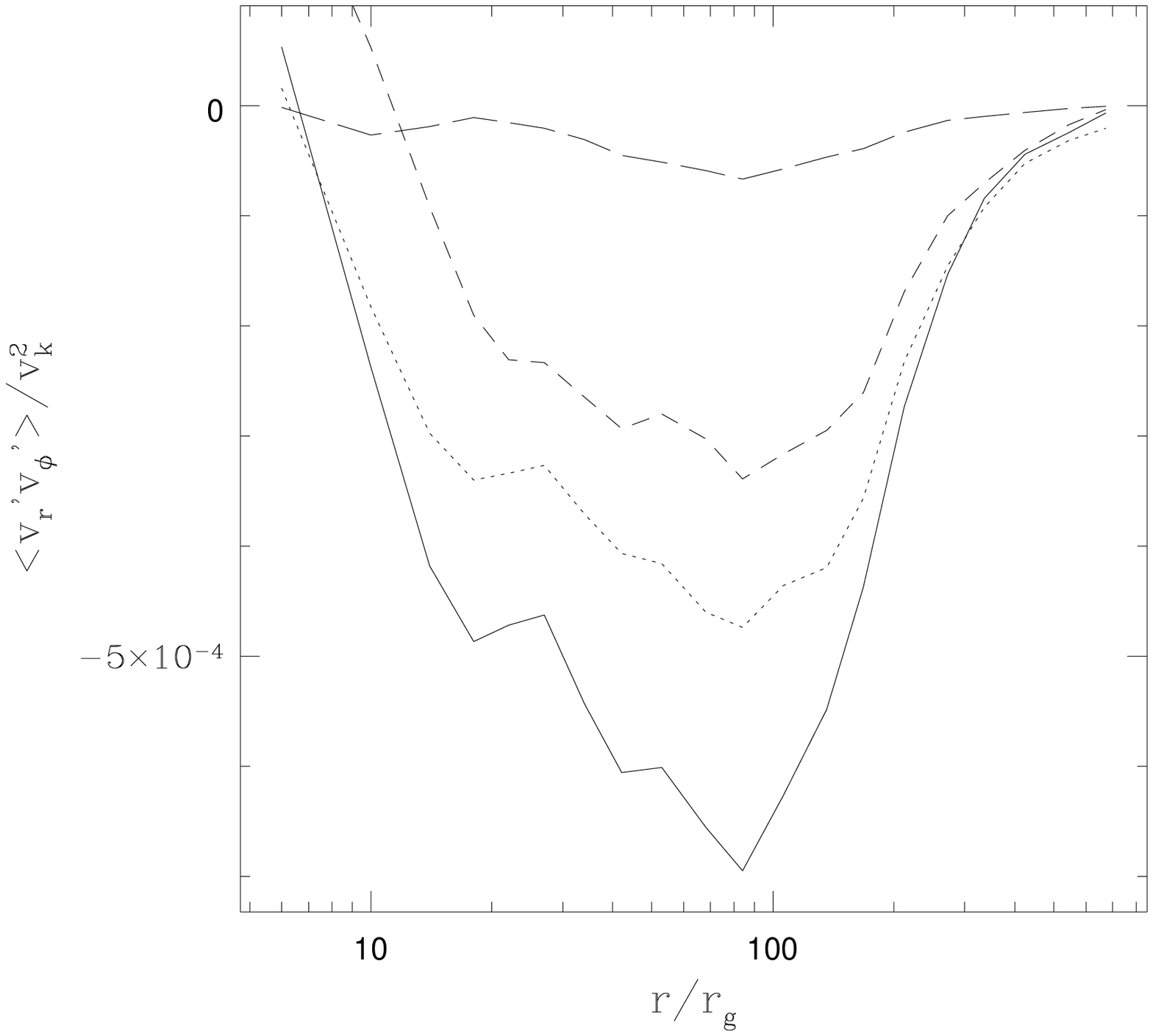}
\caption
{Radial profiles of the $(r\varphi)$-component of the Reynolds stress
tensor $\langle v'_r v'_\varphi\rangle$ for Models~A (solid line), B
(dotted line), C (dashed line) and D (long-dashed line). The stresses
have been averaged over polar angle $\theta$ and normalized to the
square of the Keplerian velocity $v_K^2$.
\label{fig5}}
\end{figure}

\placefigure{fig5}

\section{Discussion and Conclusions}

We have used numerical simulations to study three-dimensional
hydrodynamic accretion flows with inefficient radiative cooling
(ADAFs).  The simulated flows have nearly Keplerian angular momentum
on the outside, relatively weak viscosity, $\alpha=0.01$, $0.03$, and
an adiabatic index $\gamma$ equal to either $5/3$ or $4/3$.  We find
that the convective motions are nearly axisymmetric (Fig. 3).  As a
result, the basic properties of the flows are very similar to those
found in previous 2D axisymmetric simulations (Igumenshchev et
al. 1996; Igumenshchev \& Abramowicz 1999, 2000; Stone et al. 1999;
NIA).

Our most important result is that convection transports angular
momentum inward (Fig. 5).  This property of convection was identified
by Ryu \& Goodman (1992) in analytical work on thin accretion disks.
Recently, NIA showed that convection in 2D axisymmetric simulations of
ADAFs also moves angular momentum inward.  They and QG argued,
however, on the basis of previous work by Stone \& Balbus (1996), that
this result was a consequence of enforcing axisymmetry on the
accreting gas.  The present work shows that even in 3D, where the flow
is not constrained to be axisymmetric, convective motions do transport
angular momentum inward.  In fact, we find that 2D and 3D simulations
show good quantitative agreement in the estimate of the
($r\varphi$)-component of the Reynolds stress tensor, and in the
radial dependences of various fluid variables.

In the models, the turbulent pressure due to convection,
$P_{turb}=\langle\rho v'^2\rangle /3$, where $v'$ is the velocity
fluctuation, is $\la 10^{-3}$ of the gas pressure $P$.  Therefore,
convection does not have a direct dynamical effect on the flow.
However, as argued by NIA and QG, convection can have a very strong
indirect effect on the structure of the flow if it causes inward
transport of angular momentum.  The accretion flow can then achieve a
critically balanced state in which the convective transport almost
completely cancels the outward transport of angular momentum by
viscosity.  This leads to a flat radial density profile $\rho\propto
r^{-1/2}$.  Since the amount of transport possible with convection is
limited, the viscosity parameter $\alpha$ must be less than a critical
value $\alpha_{crit}\la0.1$ for the new structure to be realized
(NIA).  The value of $\alpha$ in an ADAF is not known at present.
Model-fitting to the observed light curve and spectra of soft X-ray
transients suggests that $\alpha\sim0.2-0.3$ (Esin, McClintock \&
Narayan 1997).  Numerical MHD simulations in 3D should provide an
independent estimate of $\alpha$ (assuming that viscosity in ADAFs is
produced by magnetic stresses).  If such computations confirm that
$\alpha$ in ADAFs is large, then convection may turn out to have only
limited dynamical effect on these flows.  Conversely, if
$\alpha<\alpha_{crit}$, one may expect ADAFs to resemble the
analytical solutions described by NIA and QG.

Our results may be relevant also to the physics of rapidly rotating
convective stars. In such stars convection may reduce or even suppress
the outward transport of angular momentum by magnetic stresses (see
Bisnovatyi-Kogan et al. 1979).

An important property of convective accretion flows is the outward
transport of energy by convection.  We find that 2D and 3D simulations
give similar results; the effective outward luminosity due to
convection is $\sim 10^{-3}-10^{-2}\dot{M}c^2$ for the values of
$\alpha$ and $\gamma$ considered here, where $\dot{M}$ is the net mass
accretion rate of the black hole.  Very likely the energy will be
radiated on the outside of the accretion flow.  If so, the results
have important implications for the spectra and luminosities of
accreting black holes and neutron stars (Igumenshchev \& Abramowicz
2000).  Convection could also introduce variability on different
time-scales (Igumenshchev \& Abramowicz 1999).

The simulated flows presented in this paper show some interesting
dependence on the adiabatic index $\gamma$.  Models with $\gamma=4/3$
have a steeper density profile (Fig.~2) than models with $\gamma=5/3$.
In addition, Model~D ($\alpha=0.01$, $\gamma=4/3$) shows a
quasi-stationary meridional circulation (Fig.~4), whereas other models
have chaotic convective motions.  The reason for these differences is
unclear.

\acknowledgments 
This study was supported by a grant from the Royal Swedish Academy of
Sciences and by grants PHY 9507695 and AST 9820686 from the National
Science Foundation.

\clearpage

\begin{deluxetable}{lcc}
\footnotesize
\tablecaption{Parameters of the models. \label{tbl-1}}
\tablewidth{0pt}
\tablehead{
\colhead{Model} & \colhead{~~$\alpha$~~} & \colhead{~~$\gamma$~~} }
\startdata
A & $0.03$  & $5/3$ \nl
B & $0.03$  & $4/3$ \nl
C & $0.01$  & $5/3$ \nl
D & $0.01$  & $4/3$ \nl
\enddata
\end{deluxetable}

\clearpage

\clearpage

\clearpage

\clearpage

\clearpage

\clearpage

\begin{references}
\reference{} Bisnovatyi-Kogan, G. S., Blinnikov, S. I., Kostyuk, N. D.,
    \& Fedorova, A. V. 1979, Soviet Astr., 23, 432
\reference{} Colella, P., \& Woodward, P. R. 1984, J. Comput. Phys., 54, 174
\reference{} Esin, A. A., McClintock, J. E., \& Narayan, R. 1997,
    \apj, 489, 865
\reference{} Igumenshchev, I.V., Chen, X., \& Abramowicz, M.A. 1996, \mnras,
    278, 236
\reference{} Igumenshchev, I.V., \& Abramowicz, M.A. 1999, \mnras, 303, 309
\reference{} Igumenshchev, I.V., \& Abramowicz, M.A. 2000, \apj, submitted
    (astro-ph/0003397)
\reference{} Narayan, R., \& Yi, I. 1994, \apj, 428, L13
\reference{} Narayan, R., \& Yi, I. 1995, \apj, 444, 231
\reference{} Narayan, R., Igumenshchev, I.V., \& Abramowicz, M.A. 2000,
    \apj, in press (astro-ph/9912449, NIA)
\reference{} Quataert, E. \& Gruzinov, A. 2000, \apj, in press
    (astro-ph/9912440, QG)
\reference{} Ryu, D., \& Goodman, J. 1992, ApJ, 388, 438
\reference{} Stone, J.M., Pringle, J.E., \& Begelman, M.C. 1999, \mnras,
    310, 1002
\reference{} Stone, J.M., \& Balbus, S. A. 1996, \apj, 464, 364
\end{references}
\end{document}